\documentclass[aps,pra,twocolumn,a4paper,superscriptaddress]{revtex4}

\usepackage[brazil]{babel}
\usepackage[latin1]{inputenc}

\usepackage{amsmath}
\usepackage{amssymb}
\usepackage{dcolumn}
\usepackage{graphicx} 

\begin{document}

\title{Introdução à Criptografia Quântica \\
       \small{(Introduction to Quantum Cryptography)}}

\author{Gustavo Rigolin}
\email{rigolin@ifi.unicamp.br}
\affiliation{Departamento de Raios C\'osmicos e Cronologia, Instituto de
F\'{\i}sica Gleb Wataghin, Universidade Estadual de Campinas, C.P. 6165,
cep 13084-971, Campinas, S\~ao Paulo, Brasil}
\author{Andrés Anibal Rieznik}
\email{anibal@ifi.unicamp.br}
\affiliation{Centro de Pesquisa em Ótica e Fotônica, Instituto de F\'{\i}sica
Gleb Wataghin \\ Universidade Estadual de Campinas, cep 13083-970, Campinas,
S\~ao Paulo, Brasil, e \\ PADTEC, Rodovia Campinas-Mogi-Mirim (SP 340) Km 118.5,
cep 13086-902, Campinas, S\~ao Paulo, Brasil}

\begin{abstract}
Apresentamos de maneira detalhada os quatro protocolos de distribuição de chaves  que
fundaram a importante área da criptografia quântica, numa linguagem acessível a alunos de
graduação em Física. Começamos pelo protocolo BB84, o qual se utiliza de estados de
polarização de fótons para transmitir chaves criptográficas. Em seguida, apresentamos o
protocolo E91, que faz uso de singletos para gerar uma seqüência de números aleatórios.
Finalizamos este artigo apresentando o protocolo BBM92 e o B92, os quais podem ser vistos
como simplificações dos dois primeiros protocolos. \\

We show in details the four quantum key distribution protocols which initiated the
important field of quantum cryptography, using an accessible language for undergraduate
students. We begin presenting the BB84 protocol, which uses polarization states of
photons in order to transmit cryptographic keys. Thereupon we show the E91 protocol,
whose security is based on the use of singlet states to generate a random sequence of bits.
We end the paper with the BBM92 and the B92 protocol. These last two protocols can be seen
as simplified versions of the first two.
\end{abstract}

\maketitle

\section{Introdução}

Desde os primórdios da civilização o homem sempre se deparou com o problema de
transmitir secretamente informações importantes. A ciência que estuda essa arte de se
comunicar confidencialmente, tendo a certeza de que somente as partes interessadas terão
acesso à informação, recebe o nome de criptografia.

Muitos dos modernos protocolos de crip\-to\-gra\-fia a\-nun\-ciam publicamente o
algoritmo utilizado pa\-ra encriptar e decriptar a mensagem. Ao anunciar publicamente
este procedimento, permitimos a todos, inclusive quem desejamos que não tenha acesso à
mensagem, conhecer o modo de deixá-la secreta. A segurança desses protocolos se baseia
apenas em uma longa seqüência de números aleatórios que o emissor (Alice) e o receptor (Bob)
da mensagem devem compartilhar em segredo. Ninguém mais pode conhecer esses números.
 Ou seja, o sucesso desses protocolos depende exclusivamente da capacidade de os
 envolvidos na comunicação serem capazes de compartilhar essa seqüência de números
 aleatórios, também conhecida como chave criptográfica, certificando-se de que ninguém
 mais consiga ter acesso a ela.

Para compartilhar essa chave, Alice e Bob usam um canal clássico
de comunicação. Por mais seguro que ele seja, em princípio ele
pode ser monitorado por algum agente externo (Eva) sem que Alice e
Bob percebam. Eva pode obter a chave, sem Alice e Bob notarem,
pois qualquer informação clássica pode ser clonada. Eva pode, por
exemplo, interceptar a chave enviada por Alice a Bob e, em
seguida, reenviá-la a ele. Hoje em dia, no entanto, para contornar
este problema utilizamos os, assim chamados, protocolos de chave
pública (amplamente utilizados nas transações financeiras via
internet). Sua segurança, porém, não é matematicamente provada e
desabaria perante o aparecimento de computadores quânticos. Na
seção seguinte nos detemos um pouco mais neste aspecto.

Agora, se Alice e Bob usarem um canal quântico de comunicação,
eles terão certeza de que a transmissão da chave foi realizada com
segurança total, ou de que ela foi interceptada por Eva. Essa
segurança é baseada nas leis da Mecânica Quântica e, desde que
aceitemos que ela é uma teoria completa no sentido de Bohr
\cite{bohr,epr}, não há meio de se burlar essa segurança.

O primeiro protocolo de criptografia quântica, ou mais corretamente, protocolo de
distribuição de chaves quânticas, foi proposto por Bennett e Brassard, no ano de
1984 \cite{bb84}. Ele também é conhecido como protocolo BB84. É usual, entre os
criptólogos, nomear um protocolo de criptografia usando-se as iniciais dos nomes
 dos autores que o criaram mais o ano de sua invenção. A transmissão da chave é
  feita enviando-se fótons que podem ser preparados em quatro estados de
  polarização. Os fótons, neste protocolo, não estão emaranhados. Entretanto,
   Artur K. Ekert criou um protocolo (E91) \cite{ekert} que faz uso do estado
    de Bell $|\Psi^{-}\rangle = (1/\sqrt{2})(|01\rangle - |10\rangle)$ para
    transmitir chaves quânticas. Sua segurança está baseada na impossibilidade
    de violação da desigualdade de Clauser-Horne-Shimony-Holt (CHSH) \cite{chsh}.
     Em 1992, Bennett, Brassard e Mermin \cite{bbm92} simplificaram o protocolo E91
      criando o protocolo BBM92 e provaram de um modo muito simples e profundo a
       impossibilidade de as chaves serem conhecidas por outra pessoa sem Alice e
        Bob perceberem. Também em 1992, Bennett \cite{b92} criou o protocolo B92,
         no qual apenas dois estados de polarização de fótons são utilizados para
          se transmitir seguramente uma chave criptográfica.

Neste artigo discutimos em detalhes os quatro protocolos de transmissão de chaves
quânticas acima mencionados. Pretendemos apresentá-los do modo mais simples e
intuitivo possível, pois acreditamos que alguns deles já podem e devem ser
 ensinados durante um curso de graduação em Física. Alunos que já tenham ou
 estejam estudando Mecânica Quântica (MQ) conseguem, sem muito esforço, entendê-los.
 Acreditamos também  que estes protocolos possam vir a ser ferramentas muito úteis
 até para se ensinar MQ, pois eles representam aplicações práticas e importantes de
  conceitos inerentes ao mundo quântico.

\section{BB84}

Uma das particularidades da Criptografia Quântica (CQ) está no
fato de que a melhor forma de se começar a entendê-la e estudá-la
consiste na leitura do primeiro artigo dedicado a esse assunto
\cite{bb84}.  Em ciência isto é um fato raro. Não é comum
recomendar a um aluno iniciando-se em alguma área do conhecimento
a leitura dos artigos que a fundaram. Geralmente, após a aparição
desses artigos, outros mais simples e pedagógicos são publicados,
os quais são mais adequados para um aprendiz.  Felizmente isso não
ocorre com a CQ. De fato, a simplicidade da Ref. \cite{bb84} a
torna não somente ampla e unanimemente reconhecida como a
fundadora da CQ, como também, a nosso ver, a melhor introdução a
essa área. Ela tem quatro páginas e está escrita numa linguagem
tão clara, precisa e direta que qualquer aluno que tenha feito um
curso introdutório de Mecânica Quântica pode entendê-la sem muito
esforço. A Ref. \cite{bb84} se enquadra na honrosa categoria de
trabalhos científicos que podem ser lidos ao sofá, em 15 minutos,
desfrutando-se de um bom café. No restante desta seção vamos fazer
um resumo deste artigo, enfatizando aspectos que serão
fundamentais para o entendimento dos outros protocolos de CQ
apresentados nas seções seguintes. Para os leitores mais
interessados, recomendamos com entusiasmo a leitura do trabalho
original de Bennett e Brassard, o qual pode ser gratuitamente
copiado a partir da página pessoal de Charles H. Bennett:
www.research.ibm.com/people/b/bennetc/chbbib.\-htm.

Antes de iniciar a exposição do protocolo BB84, vale a pena dizer
que este protocolo é usado em todos os sistemas bem-sucedidos de
CQ instalados até hoje e, mais ainda, ele é o único oferecido por
duas companhias especializadas em segurança de transmissão de
dados. Assim, mesmo sendo o primeiro protocolo proposto na
literatura, ele ainda é, apesar das muitas alternativas de CQ
apresentadas \textit{a posteriori}, aquele de maior importância
prática e comercial.

A Ref. \cite{bb84} apresenta, pela primeira vez, a idéia de que a
Mecânica Quântica (MQ) pode ser utilizada para alcançar uma das
principais metas da criptografia, i.e., a distribuição segura de
uma chave criptográfica (seqüência de números aleatórios) entre
duas partes (Alice e Bob) que inicialmente não compartilham
nenhuma informação secreta. Para isso, Alice e Bob devem dispor
não só de um canal quântico, mas também de algum canal clássico de
comunicação. Este último pode ser monitorado passiva mas não
ativamente por um agente externo (Eva). Por meio dessa chave,
Alice e Bob podem com absoluta certeza se comunicar com segurança.
A garantia da distribuição segura de chaves por meio da CQ se
sustenta na validade da MQ tal qual a conhecemos. Em contraste, a
criptografia de chave pública é considerada segura devido a um
suposto grau de complexidade matemática inerente ao algoritmo de
decodificação necessário para recuperar a mensagem criptografada
se não conhecemos a chave privada. No entanto, esse resultado
nunca foi matematicamente provado e não há nada que impeça a
criação de um algoritmo (quem sabe ele já não esteja nas mãos de
alguma agência de inteligência governamental) que possa facilmente
decodificar, por meio de computadores convencionais, mensagens
secretas oriundas de protocolos de chaves públicas. Pior ainda, a
segurança da criptografia de chave pública tradicional desabaria
perante o aparecimento de computadores quânticos, o que não
aconteceria com sistemas de distribuição de chaves por CQ.

Na Introdução da Ref. \cite{bb84}, após uma breve digressão sobre sistemas
criptográficos tradicionais, os autores explicitam claramente as novidades
 que serão apresentadas no artigo: como utilizar a MQ para (1) criar
 protocolos de transmissão segura de chaves criptográficas e
 (2) ``jogar cara-ou-coroa"  sem possibilidade de enganar o oponente.
 Após a Seção II, onde os autores apresentam o formalismo a ser utilizado,
 na Seção III discute-se o protocolo para distribuição de chaves e na Seção
  IV o protocolo para jogar cara-ou-coroa sem trapacear. Aqui expomos apenas
   o protocolo de transmissão de chave criptográfica, i. e., o famoso protocolo BB84.

\begin{figure}[!ht]
\begin{center}
\includegraphics[width=2in]{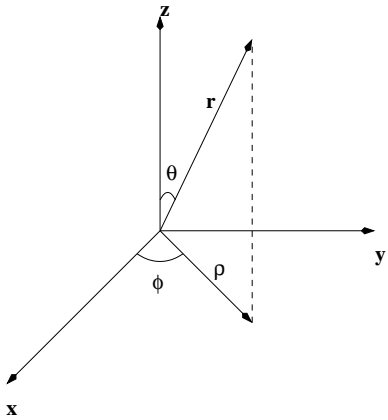}
\caption{\label{esferica}\small{Coordenadas esféricas. O ângulo polar
 $\theta$ varia de $0$ a $\pi$ e o ângulo azimutal $\phi$ de $0$ a $2\pi$.
 Aqui, o vetor $\mathbf{r}$, de módulo $r$, tem projeção no plano $xy$ dada
  por $\rho = r \sin \theta $. As coordenadas cartesianas se relacionam com
   as coordenadas esféricas pela seguinte equação:
   $\mathbf{r} = x\;\hat{\mathbf{x}} + y\;\hat{\mathbf{y}} + z\;\hat{\mathbf{z}}
    =r\sin\theta\cos\phi \;\hat{\mathbf{x}} + r\sin\theta\sin\phi \; \hat{\mathbf{y}}
    + r\cos\theta \; \hat{\mathbf{z}}$.}}
\end{center}
\end{figure}

Este protocolo utiliza-se de sistemas quânticos de dois níveis. Assim,
 os estados $|0\rangle$ e $|1\rangle$ representam fótons linearmente
 polarizados em direções ortogonais. Por exemplo, os estados
 $|0\rangle$ e $|1\rangle$ podem representar fótons que se
  propagam na direção $z$ com campos elétricos oscilando no
  plano $xy$. As direções de polarização são representadas por
  vetores unitários. Usando coordenadas esféricas, de acordo com
   a notação definida na Fig. \ref{esferica}, precisamos de dois
    parâmetros (ângulos) para especificar uma direção de
    polarização.

\begin{figure}[!ht]
\begin{center}
\includegraphics[width=2.75in]{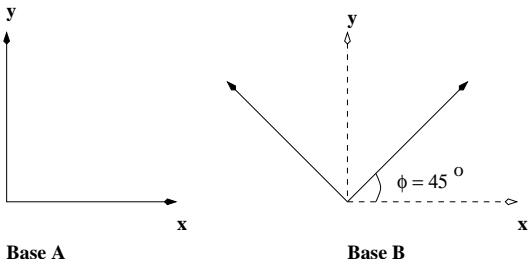}
\caption{\label{figbase}\small{Representação das bases A e B. O
eixo $z$ não está desenhado pois temos polarizações pertencentes
ao plano $xy$.}}
\end{center}
\end{figure}

Alice e Bob devem primeiramente escolher duas bases que serão
 utilizadas para a transmissão e recepção dos fótons. Cada base
  é composta por dois estados ortogonais de polarização. Eles podem
  escolher, por exemplo, polarizações contidas no plano $xy$ ($\theta=\pi/2$).
    Tomando $\phi=0$ e $\phi=\pi/2$ definimos as direções de polarização de
     uma das bases (base A). Usando $\phi=\pi/4$ e $\phi=3\pi/4$ obtemos a
      outra (base B). Veja Fig. \ref{figbase}.

O estado de polarização de qualquer fóton pode ser representado
como uma combinação linear de dois estados ortogonais de polarização.
Dessa forma, por meio dos estados que formam a base A ou a base B,
podemos representar qualquer estado de polarização de um fóton.

Alice e Bob também devem combinar previamente quais estados
ortogonais de cada uma das bases representam o bit $0$ e o bit $1$.
 Isso pode ser feito via um canal tradicional (clássico) de
  comunicação. No nosso exemplo, utilizamos os fótons polarizados
   na direção $\phi=0$ ou $\phi=\pi/4$ para representar o bit $0$
   ($|0\rangle_{A}$ e $|0\rangle_{B}$) e aqueles com polarização
   na direção $\phi=\pi/2$ ou $\phi=3\pi/4$ representando o bit
   $1$ ($|1\rangle_A$ e $|1\rangle_B$). Nesta notação, o subíndice
   em cada ket indica se temos fótons polarizados nos autoestados
   da base A ou B. Note que
    $|0\rangle_B = (1/\sqrt{2})(|0\rangle_{A} + |1\rangle_A)$
    e $|1\rangle_{B}=(1/\sqrt{2})(|0\rangle_{A} - |1\rangle_{A})$.

Alice, para transmitir a chave, procede da seguinte forma.
Primeiro ela escolhe qual se\-qüên\-cia a\-le\-a\-tó\-ria de
 bits enviará a Bob (vamos usar, por exemplo, $001111...$).
 Depois, qual a base utilizada para transmitir cada bit. Ela
  pode transmitir os dois primeiros bits utilizando-se da Base A,
   os três bits seguintes utilizando-se da Base B, o bit seguinte
    utilizando-se novamente da Base A, e assim por diante. Dessa
     forma, ela estaria enviando a Bob uma seqüência de fótons
     representados pelos seguintes kets:
     $|0\rangle_{A}, |0\rangle_{A}, |1\rangle_{B}, |1\rangle_{B},
     |1\rangle_{B}, |1\rangle_{A}$, etc. Bob, por sua vez, deve
      escolher apenas qual base ele irá utilizar para detectar
      cada fóton. Ele oscila entre as bases A e B aleatoriamente.

Após a transmissão e detecção dos fótons, Alice e Bob revelam
 publicamente quais bases utilizaram para enviar e detectar
  cada fóton, respectivamente. Mas Alice não revela se enviou
  $0$s ou $1$s e Bob não revela o resultado de suas medidas.
  Apenas as bases utilizadas (base A ou base B) são publicamente
   reveladas. A seguir, eles consideram apenas os resultados nos
    quais ambos utilizaram a mesma base, descartando todos os
     demais. Agora eles revelam publicamente uma parte destes
     resultados (metade, ou um terço, por exemplo).
     Se Eva não monitorou a transmissão, os resultados
     revelados por Bob e Alice devem coincidir; mas se
     ela a monitorou, a probabilidade de que todos os dados
     públicos coincidam é praticamente nula (provamos isso um
      pouco mais à frente). Se os dados revelados publicamente
      coincidirem, isso será uma prova de que Eva não monitorou a
       transmissão e eles podem usar o restante dos dados como a
       chave. (Por restante dos dados entendemos aqueles nos quais
       ambos usaram a mesma base para enviar e medir os fótons.)
       E aqui termina o protocolo.

Se Eva monitorou os dados, a parte da informação revelada
publicamente por Alice e Bob não irá coincidir ou, mais
rigorosamente, a probabilidade de que elas coincidam é
praticamente nula. A prova deste fato é como segue. Para
simplificar a demonstração e sem perda de generalidade, supomos
que Alice, Bob e Eva utilizam metade das vezes a Base A e metade
das vezes a Base B, Alice para transmitir e Eva e Bob para
detectar os fótons. Se Alice e Bob utilizam a mesma base, a
probabilidade de  Eva usar a mesma base vale $0.5$ (se Alice e Bob
utilizaram a Base A, por exemplo, a probabilidade de Eva também
ter utilizado essa base é $0.5$). Agora, se Eva utiliza para
monitorar os fótons a outra base, a probabilidade de Bob medir
corretamente o valor do bit transmitido é de apenas $0.5$ e não
$1$, como deveria ser se não tivéssemos um espião ou se Eva
tivesse optado pela base correta. Formalmente, suponhamos que
Alice enviou o fóton representando o bit $1$, na Base A
($|1\rangle_{A}$ ) e Bob corretamente mediu na base A, porém Eva
mediu o fóton, antes de ele chegar a Bob, na base B. Procedendo
dessa forma, Eva terá colapsado o estado de polarização dos fótons
em um dos autovetores da base por ela utilizada, i.e.,
$|0\rangle_{B} = (1/\sqrt{2})(|0\rangle_{A} + |1\rangle_{A})$ ou
$|1\rangle_{B} = (1/\sqrt{2})(|0\rangle_{A} - |1\rangle_{A})$.
Assim, quando Bob realizar sua medida, a chance de ele medir,
$|1\rangle_{A}$ é de apenas $(1/\sqrt{2})^{2} = 0.5$, independente
do resultado obtido por Eva. O fato de Eva escolher a base errada
implica, para um evento,  uma probabilidade igual a $0.5$ de Bob
detectar o valor correto para o bit transmitido por Alice. Para
uma chave muito grande, a probabilidade de Bob detectar todos os
bits corretamente, com Eva interferindo, tende a zero ou, mais
rigorosamente, a $(0.5)^{N}$, onde $N$ é o número de vezes que Eva
usou a base errada.

Vale a pena lembrar que estados quânticos arbitrários não
podem ser clonados. Isso foi demonstrado independentemente por
Wootters e Zureck \cite{clone} e por Dieks \cite{fulano}.  Isso
garante que Eva não pode simplesmente duplicar o estado quântico
 dos fótons enviados por Alice, medir um deles e enviar a Bob o outro.
  Isso possibilitaria a Eva detectar a polarização correta dos fótons
  transmitidos por Alice sem ser descoberta, tornando o protocolo BB84 inseguro.

Para melhor entender todas as etapas do protocolo,
a Tab. \ref{bb84tabela1} simula um exemplo de transmissão de
chave quântica. Consideramos uma situação bem geral, na qual
alguns fótons podem se perder durante a transmissão, de forma
que Bob não os recebe.

Finalizamos esta seção contando dois fatos relacionados ao
nascimento desta primeira proposta de CQ. Acreditamos que estas duas
histórias são de interesse para estudantes de graduação em Física.
Ambas foram extraídas do livro \textit{The Code Book},  de Simon Singh
\cite{codebook}. Elas revelam bastante bem as angústias e momentos de
tensão pelas quais muitos Físicos passam durante alguns (para não dizer vários)
 momentos de suas vidas profissionais.

\begin{widetext}
\begin{center}
\begin{table}[!ht]
\caption{\label{bb84tabela1} As cinco primeiras linhas
correspondem à transmissão quântica. As outras cinco, à discussão
pública entre Alice e Bob. A última representa a chave
compartilhada por eles.}
\begin{ruledtabular}
\begin{tabular}{lllllllllllll}
Seqüência de bits de Alice & 0 & 1 & 1 & 0 & 1 & 1 & 0 & 0 & 1 & 0 & 1 & 1 \\
Bases escolhidas por Alice & B & A & B & A & A & A & A & A & B & B & A & B \\
Fótons enviados por Alice & $|0\rangle_{B}$ & $|1\rangle_{A}$ & $|1\rangle_{B}$
& $|0\rangle_{A}$ & $|1\rangle_{A}$ & $|1\rangle_{A}$ & $|0\rangle_{A}$
& $|0\rangle_{A}$ & $|1\rangle_{B}$ & $|0\rangle_{B}$  & $|1\rangle_{A}$
& $|1\rangle_{B}$ \\
Bases escolhidas por Bob & A & B & B & A & A & B & B & A & B & A & B & B \\
Bits recebidos por Bob & 1 & & 1 & & 1 & 0 & 0 & 0 &  & 1 & 1 & 1 \\ \hline
Bob informa fótons detectados & A & & B & & A & B & B & A & & A & B & B \\
Alice informa bases corretas & & & OK & & OK & & & OK & & & & OK \\
Informação compartilhada & & & 1 & & 1 & & & 0 & & & & 1 \\
Bob revela alguns bits da chave & & & & & 1 & & & & & & & $\,$ \\
Alice confirma estes bits & & & & & OK & & & & & & & $\,$ \\ \hline
Restante de bits é a chave & & & 1 & & & & & 0 & & & & 1
\end{tabular}
\end{ruledtabular}
\end{table}
\end{center}
\end{widetext}

A primeira delas trata-se de uma pessoa que estava à frente de seu tempo:
Stephen Wiesner, quem, em 1960, teve a idéia de utilizar a MQ de forma
parecida a utilizada hoje em CQ. Ele demonstrou a possibilidade teórica
de se fazer ``dinheiro quântico", impossível de ser falsificado graças
a um sistema de armazenamento quântico de bits. Longe de prática, a
idéia era, porém, revolucionária. Anos mais tarde Bennett e Brassard
inspiraram-se nessa idéia de ``dinheiro quântico" para criar o protocolo BB84.
 Contudo, o mais interessante dessa história consiste em a idéia de
  Wiesner ter sido absolutamente ignorada no seu tempo. O seu
  orientador pediu-lhe que abandonasse a idéia e voltasse ao ``trabalho",
   mostrando total desinteresse por ela. Conta Wiesner: ``Não obtive
    nenhum apoio do meu orientador de tese - ele não mostrou o mínimo
     interesse pela minha idéia. Mostrei-a para outras várias pessoas
     e todas fizeram uma cara de estranheza e voltaram ao que já
     estavam fazendo naquela hora". Apesar disso, Wiesner submeteu
      a sua idéia para ser publicada numa revista científica.
      O artigo foi recusado. Submeteu-o a outras três revistas,
      e outras três vezes ele foi recusado. Desiludido, e consciente
      do grande interesse de Bennett por assuntos mais amplos,
      Wiesner enviou o seu rejeitado artigo a ele. Bennett ficou
      imediatamente fascinado pela idéia, mostrando-a para Brassard.
       Alguns anos depois, os dois juntos, inspirados na idéia de
        utilizar a MQ como proposto por Wiesner, inventaram o hoje
        em dia reconhecido e aclamado campo da CQ.

A segunda história pitoresca refere-se ao exato momento no qual
Bennett e Brassard inventaram o protocolo BB84. Em 1984 fazia já
algum tempo que ambos vinham tentando achar uma solução para o
problema da distribuição de chaves, num cenário futurístico onde a
computação quântica inviabilizara os atuais métodos de
criptografia de chave pública. Um dia, quando estavam esperando o
trem que levaria Brassard a seu lar, em Montreal, desde os
laboratórios Thomas. J. Watson, da IBM, onde Bennett trabalhava, a
solução para o problema surgiu. Esperando o trem na estação
Croton-Harmon, conversando descontraída e informalmente, num
momento de \textit{eureka}, eles tiveram a brilhante idéia que
levou ao protocolo BB84. Como afirma Simon Singh em \textit{The
Code Book}, se o trem tivesse chegado apenas alguns minutos antes
eles teriam se despedido sem fazer nenhum progresso no problema da
distribuição de chaves.

\section{E91}

Alice e Bob dispõem, agora, de um canal quântico que emite
singletos: $|\Psi^{-}\rangle = (1/\sqrt{2})(|01\rangle -
|10\rangle)$\footnote{Vale a pena observar que o estado
$|\Psi^{-}\rangle$ é um estado emaranhado e não-local. Ele não
pode ser escrito como um produto tensorial de um único estado
pertencente a Alice e de um outro pertencente a Bob.}.
 Alice recebe um
dos constituintes do singleto enquanto Bob recebe o
 outro. Vamos supor, sem perder em generalidade, que as partículas
 viajam até Alice e Bob ao longo da direção $z$. Ao receberem-nas,
 Alice e Bob medem o spin de suas partículas ao longo da direção
 $\mathbf{a}_{i}$ e $\mathbf{b}_{j}$, respectivamente. O vetor
 $\mathbf{a}_{i}$ ($\mathbf{b}_{j}$) é unitário e caracterizado
  pelos ângulos polar $\theta^{a}_{i}$ ($\theta^{b}_{j}$) e
  azimutal $\varphi^{a}_{i}$ ($\varphi^{b}_{j}$). Veja a
  Fig. \ref{esferica}. Tanto Alice quanto Bob orientam,
  aleatoriamente para cada medida de spin, seus detectores
  ao longo de três vetores contidos no plano $xy$, i. e.
  $\theta^{a}_{i} = \theta^{b}_{j} = \pi/2$. Os ângulos
  azimutais que caracterizam estes vetores são: $\varphi^{a}_{1} = 0$,
   $\varphi^{a}_{2} = \pi/4$ e $\varphi^{a}_{3} = \pi/2$ para Alice,
    e $\varphi^{b}_{1} = \pi/4$, $\varphi^{b}_{2} = \pi/2$ e $\varphi^{b}_{3}
    = 3\pi/4$ para Bob.

A partir destes vetores, podemos definir o coeficiente de
 correlação de medidas de spin (polarização) ao longo das
  direções $\mathbf{a}_{i}$ e $\mathbf{b}_{j}$ como sendo
\begin{eqnarray}
E(\mathbf{a}_{i}, \mathbf{b}_{j}) & = & P_{00}(\mathbf{a}_{i},\mathbf{b}_{j})
+  P_{11}(\mathbf{a}_{i},\mathbf{b}_{j}) \nonumber \\
& & -  P_{01}(\mathbf{a}_{i},\mathbf{b}_{j}) -
P_{10}(\mathbf{a}_{i},\mathbf{b}_{j}). \label{cor}
\end{eqnarray}
Aqui $P_{00}(\mathbf{a}_{i},\mathbf{b}_{j}), P_{11}(\mathbf{a}_{i},\mathbf{b}_{j}),
 P_{01}(\mathbf{a}_{i},\mathbf{b}_{j})$ e $P_{10}(\mathbf{a}_{i},\mathbf{b}_{j})$
 re\-pre\-sen\-tam a probabilidade de obtermos os resultados $(+1,+1)$, $(-1,-1)$,
  $(+1,-1)$ e $(-1,+1)$ ao longo das direções $\mathbf{a}_{i}$ e $\mathbf{b}_{j}$,
  respectivamente. Atribuímos o valor $1$ para uma medida do estado
  $|0\rangle$ e valor $-1$ para uma medida do estado $|1\rangle$. Assim,
  o coeficiente de correlação nada mais é do que a probabilidade de Alice
   e Bob medirem o mesmo valor de spin menos a probabilidade de obterem
   valores diferentes.

Para um estado puro, a Eq.~(\ref{cor}) pode ser escrita da seguinte forma,
\begin{equation}
E(\mathbf{a}_{i}, \mathbf{b}_{j}) =
\langle \Psi^{-} | \sigma^{A}_{\mathbf{a}_{i}} \otimes \sigma^{B}_{\mathbf{b}_{j}}
 | \Psi^{-} \rangle, \label{forma}
\end{equation}
onde $\sigma^{A}_{\mathbf{a}_{i}} = \mathbf{a}_{i} \cdot \mathbf{\sigma}^{A}$
e $\sigma^{B}_{\mathbf{b}_{j}} = \mathbf{b}_{j} \cdot \mathbf{\sigma}^{B}$,
 $\mathbf{\sigma}^{A}= (\sigma^{A}_{x},\sigma^{A}_{y},\sigma^{A}_{z})$
 e $\mathbf{\sigma}^{B}= (\sigma^{B}_{x},\sigma^{B}_{y},\sigma^{B}_{z})$
  e o ponto representa o produto escalar. Para vermos isso basta lembrar
  que qualquer estado puro de dois qbits pode ser escrito como
  $|\Psi^{-}\rangle = a|0\rangle_{\mathbf{a}_{i}}|0\rangle_{\mathbf{b}_{j}}
   + b|0\rangle_{\mathbf{a}_{i}}|1\rangle_{\mathbf{b}_{j}} +
    c|1\rangle_{\mathbf{a}_{i}}|0\rangle_{\mathbf{b}_{j}} +
    d|1\rangle_{\mathbf{a}_{i}}|1\rangle_{\mathbf{b}_{j}}$,
     com $a,b,c$ e $d$ complexos, $|0(1)\rangle_{\mathbf{a}_{i}}$
     autovetor de $\sigma^{A}_{\mathbf{a}_{i}}$  e
     $|0(1)\rangle_{\mathbf{b}_{j}}$ autovetor de
     $\sigma^{B}_{\mathbf{b}_{j}}$. Substituindo essa expansão de
     $|\Psi^{-}\rangle$ na Eq.~(\ref{forma}) obtemos
\begin{equation}
E(\mathbf{a}_{i}, \mathbf{b}_{j}) = |a|^{2} + |d|^{2} - |b|^{2} - |c|^{2}.
\end{equation}
Agora, como $|a|^{2} = P_{00}(\mathbf{a}_{i},\mathbf{b}_{j})$,
$|d|^{2} = P_{11}(\mathbf{a}_{i},\mathbf{b}_{j})$,
$|b|^{2} = P_{01}(\mathbf{a}_{i},\mathbf{b}_{j})$
e $|c|^{2}$ $=$ $P_{10}(\mathbf{a}_{i},\mathbf{b}_{j})$,
recuperamos a Eq.~(\ref{cor}) a partir de (\ref{forma}).

Sendo as componentes cartesianas dos vetores
$\mathbf{a}_{i} = (a_{x},a_{y},a_{z})$ e $\mathbf{b}_{j} =(b_{x},b_{y},b_{z})$
temos que
\begin{eqnarray}
\sigma^{A}_{\mathbf{a}_{i}} \otimes \sigma^{B}_{\mathbf{b}_{j}} & = &
 a_{x}b_{x}\sigma^{A}_{x}\sigma^{B}_{x} +
  a_{x}b_{y}\sigma^{A}_{x}\sigma^{B}_{y} \nonumber \\
& & + a_{x}b_{z}\sigma^{A}_{x}\sigma^{B}_{z} +
a_{y}b_{x}\sigma^{A}_{y}\sigma^{B}_{x} \nonumber \\
& & + a_{y}b_{y}\sigma^{A}_{y}\sigma^{B}_{y} +
a_{y}b_{z}\sigma^{A}_{y}\sigma^{B}_{z} \nonumber \\
 & & + a_{z}b_{x}\sigma^{A}_{z}\sigma^{B}_{x} +
  a_{z}b_{y}\sigma^{A}_{z}\sigma^{B}_{y} \nonumber \\
& & + a_{z}b_{z}\sigma^{A}_{z}\sigma^{B}_{z}. \label{prod}
\end{eqnarray}

Já que $\sigma^{A}_{\mathbf{a}_{i}} \otimes \sigma^{B}_{\mathbf{b}_{j}}$
é um observável (sua média tem que ser real) e $\sigma_{y}|0\rangle
= \mathrm{i}|0\rangle$ e $\sigma_{y}|1\rangle = - \mathrm{i}|1\rangle$,
somente termos com um número par de $\sigma_{y}$'s na Eq.~(\ref{prod})
são relevantes no cálculo de $E(\mathbf{a}_{i}, \mathbf{b}_{j})$.
Além disso, como as aplicações de $\sigma_{x}$ e $\sigma_{y}$
em $|0(1)\rangle$ produzem estados ortogonais e a aplicação de
 $\sigma_{z}$ produz apenas uma fase global no estado em que
 ele atua, termos que possuem um número ímpar de $\sigma_{z}$'s
 se anulam. Dessa forma, os únicos termos da Eq.~(\ref{prod})
 contribuindo no cálculo de $E(\mathbf{a}_{i}, \mathbf{b}_{j})$ são
\begin{eqnarray}
E(\mathbf{a}_{i}, \mathbf{b}_{j}) & = &
\langle \Psi^{-} |  a_{x}b_{x}\sigma^{A}_{x}\sigma^{B}_{x}
+ a_{y}b_{y}\sigma^{A}_{y}\sigma^{B}_{y} |\Psi^{-}\rangle \nonumber \\
& & + \langle \Psi^{-}| a_{z}b_{z}\sigma^{A}_{z}\sigma^{B}_{z} | \Psi^{-}
\rangle \nonumber
\end{eqnarray}
\begin{eqnarray}
& = & - \frac{a_{x}b_{x}}{2}\left(\langle 01 | \sigma^{A}_{x}\sigma^{B}_{x}  | 10 \rangle
 +  \langle 10 | \sigma^{A}_{x}\sigma^{B}_{x}  | 01 \rangle \right) \nonumber \\
& & - \frac{a_{y}b_{y}}{2}\left(\langle 01 | \sigma^{A}_{y}\sigma^{B}_{y}  | 10 \rangle
 +  \langle 10 | \sigma^{A}_{y}\sigma^{B}_{y}  | 01 \rangle \right) \nonumber \\
& & + \frac{a_{z}b_{z}}{2}\left(\langle 01 | \sigma^{A}_{z}\sigma^{B}_{z}  | 01 \rangle
+  \langle 10 | \sigma^{A}_{z}\sigma^{B}_{z}  | 10 \rangle \right) \nonumber \\
& = & -(a_{x}b_{x} +a_{y}b_{y} + a_{z}b_{z}) \nonumber \\
& = & - \mathbf{a}_{i} \cdot \mathbf{b}_{j}.
\end{eqnarray}
Como era de se esperar, se Alice e Bob medem seus qbits na mesma direção,
$\mathbf{a}_{i} = \mathbf{b}_{j}$, obtemos $E(\mathbf{a}_{i}, \mathbf{a}_{i}) = -1$.
 Isto expressa, para este caso em particular, o fato de que o novo estado descrevendo
 o par de qbits sempre será $|01\rangle$ ou  $|10\rangle$, não importando a orientação
  do vetor $\mathbf{a}_{i}$.

Precisamos definir só mais uma quantidade \cite{chsh} antes de apresentarmos o
protocolo de transmissão de chave quântica \cite{ekert}:
\begin{equation}
S \equiv  E(\mathbf{a}_{1}, \mathbf{b}_{1}) - E(\mathbf{a}_{1}, \mathbf{b}_{3}) +
 E(\mathbf{a}_{3}, \mathbf{b}_{1}) + E(\mathbf{a}_{3}, \mathbf{b}_{3}).
\end{equation}
Como o ângulo formado por todos os pares de vetores que aparecem acima vale $\pi/4$,
 exceto para o par $\mathbf{a}_{1}$ e $\mathbf{b}_{3}$, o qual é de $3\pi/4$, temos
  que $E(\mathbf{a}_{1}, \mathbf{b}_{1})= - E(\mathbf{a}_{1}, \mathbf{b}_{3}) =
  E(\mathbf{a}_{3}, \mathbf{b}_{1}) = E(\mathbf{a}_{3}, \mathbf{b}_{3}) = -\sqrt{2}/2$.
  Portanto,
\begin{equation}
S =  -2\sqrt{2}. \label{chsh}
\end{equation}

Voltando ao protocolo, após Alice e Bob finalizarem as medidas nos
vários pares de qbits oriundos de singletos, eles anunciam
publicamente as orientações escolhidas para cada medida e se
detectaram ou não seus qbits. Eles descartam todas as medidas em
que pelo menos um deles não detectou nenhum qbit. Isso ocorre pois
o detector não tem eficiência um. Em seguida eles separam todas as
suas medidas em dois grupos: $1$) grupo de todas as medidas nas
quais Alice e Bob usaram orientações diferentes em seus
detectores; $2$) grupo onde ambos usaram a mesma orientação
($\{\mathbf{a}_{2}, \mathbf{b}_{1}\}$ e $\{\mathbf{a}_{3},
\mathbf{b}_{2}\}$). Feita essa triagem, Alice e Bob anunciam
publicamente os resultados obtidos para todas as medidas do grupo
$1$. A partir destes dados eles calculam $S$, cujo resultado deve
ser igual ao fornecido pela Eq.~(\ref{chsh}). Se esse resultado se
verificar, eles podem utilizar os dados do grupo $2$, os quais
estão anti-correlacionados, como chave criptográfica. Caso o valor
de $S$ não seja aquele dado pela Eq.~(\ref{chsh}), Alice e Bob
descartam todos os seus dados e recomeçam o protocolo.

A fim de provar a segurança desse protocolo, devemos calcular o valor de $S$
supondo que um terceiro sujeito, diga-se Eva, interfira na transmissão dos qbits.
Suponhamos que Eva meça os qbits de Alice e Bob numa direção $\mathbf{n}_{a}$
e $\mathbf{n}_{b}$, respectivamente. Dessa forma, ao medir o singleto Eva obtém
uma das quatro possibilidades abaixo representadas:
\begin{eqnarray}
|\Psi^{-}\rangle & \stackrel{\text{medida}}{\longrightarrow}  &
|0\rangle_{\mathbf{n}_{a}} |0\rangle_{\mathbf{n}_{b}}, \nonumber \\
|\Psi^{-}\rangle & \stackrel{\text{medida}}{\longrightarrow}   &
|0\rangle_{\mathbf{n}_{a}} |1\rangle_{\mathbf{n}_{b}}, \nonumber \\
|\Psi^{-}\rangle & \stackrel{\text{medida}}{\longrightarrow}  &
|1\rangle_{\mathbf{n}_{a}} |0\rangle_{\mathbf{n}_{b}}, \nonumber \\
|\Psi^{-}\rangle & \stackrel{\text{medida}}{\longrightarrow}  &
|1\rangle_{\mathbf{n}_{a}} |1\rangle_{\mathbf{n}_{b}},
\end{eqnarray}
onde $|\;\cdot\;\rangle_{\mathbf{n}_{a}}$ e $|\;\cdot\;\rangle_{\mathbf{n}_{b}}$
são os autoestados dos operadores $\sigma^{A}_{\mathbf{n}_{a}}$ e
$\sigma^{B}_{\mathbf{n}_{b}}$.

Vamos definir as funções
$|\alpha(\mathbf{n}_{a},\mathbf{n}_{b})|^{2}$,
$|\beta(\mathbf{n}_{a},\mathbf{n}_{b})|^{2}$,
$|\gamma(\mathbf{n}_{a},\mathbf{n}_{b})|^{2}$ e
$|\delta(\mathbf{n}_{a},\mathbf{n}_{b})|^{2}$ para representar as
probabilidades de detecção de cada uma das respectivas quatro
possibilidades acima expostas. Explicitamos a dependência das
probabilidades em termos das orientações $\mathbf{n}_{a}$ e
$\mathbf{n}_{b}$ para realçar que elas dependem da estratégia de
medida utilizada por Eva. Sendo assim, o estado misto que chega a
Alice e Bob após Eva realizar suas medidas é
\begin{eqnarray}
\zeta & = & |\alpha(\mathbf{n}_{a},\mathbf{n}_{b})|^{2} |0\rangle_{\mathbf{n}_{a}}
|0\rangle_{\mathbf{n}_{b}} \,_{\mathbf{n}_{a}}\langle 0 |\,_{\mathbf{n}_{b}}
 \langle 0| \nonumber \\
& & + |\beta(\mathbf{n}_{a},\mathbf{n}_{b})|^{2} |0\rangle_{\mathbf{n}_{a}}
 |1\rangle_{\mathbf{n}_{b}} \,_{\mathbf{n}_{a}}\langle 0 |\,_{\mathbf{n}_{b}}
 \langle 1| \nonumber \\
& & + |\gamma(\mathbf{n}_{a},\mathbf{n}_{b})|^{2} |1\rangle_{\mathbf{n}_{a}}
|0\rangle_{\mathbf{n}_{b}} \,_{\mathbf{n}_{a}}\langle 1 |\,_{\mathbf{n}_{b}}
\langle 0| \nonumber \\
& & + |\delta(\mathbf{n}_{a},\mathbf{n}_{b})|^{2} |1\rangle_{\mathbf{n}_{a}}
 |1\rangle_{\mathbf{n}_{b}} \,_{\mathbf{n}_{a}}\langle 1 |\,_{\mathbf{n}_{b}}
  \langle 1|. \nonumber \\
\label{zeta0}
\end{eqnarray}

Para um estado misto qualquer, a Eq.~(\ref{cor}) pode ser escrita como
\begin{eqnarray}
E(\mathbf{a}_{i}, \mathbf{b}_{j})  =
\text{Tr}\left[\zeta  \left( \sigma^{A}_{\mathbf{a}_{i}} \otimes
\sigma^{B}_{\mathbf{b}_{j}}\right) \right]. \label{zeta1}
\end{eqnarray}
Novamente podemos ver isso expandindo $\zeta$ na base $\{| n
\rangle_{\mathbf{a}_{i}} | m
\rangle_{\mathbf{b}_{j}}\,_{\mathbf{a}_{i}}\langle k |
\,_{\mathbf{b}_{j}}\langle l |\}$, no qual $n,m,k,l = 0,1$. Temos
no total $16$ vetores. No entanto, ao to\-mar\-mos o tra\-ço,
apenas os elementos da diagonal serão diferentes de zero. Estes,
por sua vez, são  dados por
$P_{00}(\mathbf{a}_{i},\mathbf{b}_{j}),
 P_{11}(\mathbf{a}_{i},\mathbf{b}_{j}), - P_{01}(\mathbf{a}_{i},\mathbf{b}_{j})$
 e $- P_{10}(\mathbf{a}_{i},\mathbf{b}_{j})$, mos\-tran\-do que a Eq.~(\ref{zeta1})
  é equivalente a Eq.~(\ref{cor}).

Para simplificar as contas, e sem perder em generalidade, podemos
 expandir o vetor $\mathbf{a}_{i}$ num sistema de referência $x'y'z'$,
 onde $z'$ é paralelo a $\mathbf{n}_{a}$ e o vetor $\mathbf{b}_{j}$
 num sistema $x''y''z''$ onde $z''$ é paralelo a $\mathbf{n}_{b}$.
  Fizemos esta escolha de tal forma que as relações abaixo sejam satisfeitas:
\begin{eqnarray}
\sigma^{A}_{x'} |0(1)\rangle_{\mathbf{n}_{a}} & = &
|1(0)\rangle_{\mathbf{n}_{a}}, \label{xlinha} \\
\sigma^{B}_{x''} |0(1)\rangle_{\mathbf{n}_{b}} & = &
|1(0)\rangle_{\mathbf{n}_{b}}, \\
\sigma^{A}_{y'} |0(1)\rangle_{\mathbf{n}_{a}} & = &
\mathrm{i}(-\mathrm{i})|1(0)\rangle_{\mathbf{n}_{a}}, \\
\sigma^{B}_{y''} |0(1)\rangle_{\mathbf{n}_{b}} & = &
\mathrm{i}(-\mathrm{i})|1(0)\rangle_{\mathbf{n}_{b}}, \\
\sigma^{A}_{z'} |0(1)\rangle_{\mathbf{n}_{a}} & = &
+(-)|0(1)\rangle_{\mathbf{n}_{a}}, \\
\sigma^{B}_{z''} |0(1)\rangle_{\mathbf{n}_{b}} & = &
 +(-)|0(1)\rangle_{\mathbf{n}_{b}}. \label{zduaslinha}
\end{eqnarray}
Nestes sistemas de coordenadas,
\begin{eqnarray}
\sigma^{A} & = & \sigma^{A}_{x'}\mathbf{x'} + \sigma^{A}_{y'}\mathbf{y'}
+ \sigma^{A}_{z'}\mathbf{z'}, \\
\sigma^{B} & = & \sigma^{B}_{x''}\mathbf{x''} + \sigma^{B}_{y''}\mathbf{y''}
 + \sigma^{B}_{z''}\mathbf{z''}, \\
\mathbf{a}_{i} & = & a'_{x}\mathbf{x'} +  a'_{y}\mathbf{y'}
 + a'_{z}\mathbf{z'}, \label{ai} \\
\mathbf{b}_{j} & = & b''_{x}\mathbf{x''} +  b''_{y}\mathbf{y''}
+ b''_{z}\mathbf{z''}, \\
\mathbf{n}_{a} & = & \mathbf{z'}, \\
\mathbf{n}_{b} & = & \mathbf{z''}, \label{nb}
\end{eqnarray}
onde $\mathbf{x'},\mathbf{y'}, \mathbf{z'}$ e
$\mathbf{x''},\mathbf{y''},\mathbf{z''}$ são os versores que
expandem, respectivamente, os sistemas de referência $x'y'z'$ e $x''y''z''$.

Usando as expansões anteriores podemos escrever os operadores
$\sigma^{A}_{\mathbf{a}_{i}}$ e $\sigma^{B}_{\mathbf{b}_{j}}$ da seguinte forma:
\begin{eqnarray}
\sigma^{A}_{\mathbf{a}_{i}} = \mathbf{a}_{i}\cdot \sigma^{A}  =
  a'_{x}\sigma^{A}_{x'} +  a'_{y}\sigma^{A}_{y'} + a'_{z}\sigma^{A}_{z'},
   \label{eqsiga}\\
\sigma^{B}_{\mathbf{b}_{j}} = \mathbf{b}_{i}\cdot \sigma^{A}  =
 b''_{x}\sigma^{B}_{x''} +  b''_{y}\sigma^{B}_{y''} + b''_{z}\sigma^{B}_{z''}.
  \label{eqsigb}
\end{eqnarray}

Por meio das Eqs.~(\ref{eqsiga}) e (\ref{eqsigb}) vemos que
\begin{eqnarray*}
\sigma^{A}_{\mathbf{a}_{i}} \otimes \sigma^{B}_{\mathbf{b}_{j}}
& = &  a'_{x}b''_{x}\sigma^{A}_{x'}\sigma^{B}_{x''} +
a'_{x}b''_{y}\sigma^{A}_{x'}\sigma^{B}_{y''} \nonumber \\
& & + a'_{x}b''_{z}\sigma^{A}_{x'}\sigma^{B}_{z''} +
a'_{y}b''_{x}\sigma^{A}_{y'}\sigma^{B}_{x''} \nonumber \\
& & + a'_{y}b''_{y}\sigma^{A}_{y'}\sigma^{B}_{y''} +
a'_{y}b''_{z}\sigma^{A}_{y'}\sigma^{B}_{z''}\\
 & & + a'_{z}b''_{x}\sigma^{A}_{z'}\sigma^{B}_{x''} +
  a'_{z}b''_{y}\sigma^{A}_{z'}\sigma^{B}_{y''} \nonumber \\
& & + a'_{z}b''_{z}\sigma^{A}_{z'}\sigma^{B}_{z''}.
\end{eqnarray*}
Retornando ao cálculo do coeficiente de correlação, vemos
que substituindo a Eq.~(\ref{zeta0}) em (\ref{zeta1}) temos

\begin{eqnarray*}
E(\mathbf{a}_{i}, \mathbf{b}_{j}) =
|\alpha(\mathbf{n}_{a},\!\mathbf{n}_{b})|^{2} \,_{\mathbf{n}_{a}}
\!\langle 0 |\,_{\mathbf{n}_{b}}\! \langle 0| \Sigma |0\rangle_{\mathbf{n}_{a}}
 |0\rangle_{\mathbf{n}_{b}} \\
 + |\beta(\mathbf{n}_{a},\!\mathbf{n}_{b})|^{2} \,_{\mathbf{n}_{a}}\!
 \langle 0 |\,_{\mathbf{n}_{b}}\! \langle 1| \Sigma  |0\rangle_{\mathbf{n}_{a}}
 |1\rangle_{\mathbf{n}_{b}} \\
 + |\gamma(\mathbf{n}_{a},\!\mathbf{n}_{b})|^{2}   \,_{\mathbf{n}_{a}}
 \!\langle 1 |\,_{\mathbf{n}_{b}}\! \langle 0|\Sigma |1\rangle_{\mathbf{n}_{a}}
  |0\rangle_{\mathbf{n}_{b}} \\
 + |\delta(\mathbf{n}_{a},\!\mathbf{n}_{b})|^{2}  \,_{\mathbf{n}_{a}}
 \!\langle 1 |\,_{\mathbf{n}_{b}}\! \langle 1| \Sigma |1\rangle_{\mathbf{n}_{a}}
  |1\rangle_{\mathbf{n}_{b}},
\end{eqnarray*}
onde $\Sigma = \sigma^{A}_{\mathbf{a}_{i}} \otimes \sigma^{B}_{\mathbf{b}_{j}}$.

Observando as Eqs.~(\ref{xlinha}-\ref{zduaslinha}), os únicos termos de
 $\sigma^{A}_{\mathbf{a}_{i}} \otimes \sigma^{B}_{\mathbf{b}_{j}}$ que
 contribuem no cálculo de $E(\mathbf{a}_{i}, \mathbf{b}_{j})$ são
\begin{eqnarray}
\frac{E(\mathbf{a}_{i}, \mathbf{b}_{j})}{a'_{z}b''_{z}} & = &
|\alpha(\mathbf{n}_{a},\!\mathbf{n}_{b})|^{2}
\! \,_{\mathbf{n}_{a}}\!\langle 0 |\,_{\mathbf{n}_{b}}
\! \langle 0| \Sigma_{z} |0\rangle_{\mathbf{n}_{a}} |0\rangle_{\mathbf{n}_{b}}
 \nonumber \\
& & + |\beta(\mathbf{n}_{a},\!\mathbf{n}_{b})|^{2}
\! \,_{\mathbf{n}_{a}}\!\langle 0 |\,_{\mathbf{n}_{b}}
\! \langle 1| \Sigma_{z}  |0\rangle_{\mathbf{n}_{a}} |1\rangle_{\mathbf{n}_{b}}
\nonumber \\
& & + |\gamma(\mathbf{n}_{a},\!\mathbf{n}_{b})|^{2}
\!  \,_{\mathbf{n}_{a}}\!\langle 1 |\,_{\mathbf{n}_{b}}
\! \langle 0|\Sigma_{z} |1\rangle_{\mathbf{n}_{a}} |0\rangle_{\mathbf{n}_{b}}
\nonumber \\
& & + |\delta(\mathbf{n}_{a},\!\mathbf{n}_{b})|^{2}
\!  \,_{\mathbf{n}_{a}}\!\langle 1 |\,_{\mathbf{n}_{b}}
\! \langle 1| \Sigma_{z} |1\rangle_{\mathbf{n}_{a}} |1\rangle_{\mathbf{n}_{b}}
 \nonumber \\
E(\mathbf{a}_{i}, \mathbf{b}_{j}) & = &
f(\mathbf{n}_{a},\mathbf{n}_{b})a'_{z}b''_{z}
\label{fiadamae}.
\end{eqnarray}
Na expressão anterior $\Sigma_z = \sigma^{A}_{z'}\sigma^{B}_{z''}$
e $f(\mathbf{n}_{a},\mathbf{n}_{b})$ $=$
$|\alpha(\mathbf{n}_{a},\mathbf{n}_{b})|^{2}$ $-$
$|\beta(\mathbf{n}_{a},\mathbf{n}_{b})|^{2}$ $-$
$|\gamma(\mathbf{n}_{a},\mathbf{n}_{b})|^{2}$ $+$
$|\delta(\mathbf{n}_{a},\mathbf{n}_{b})|^{2}$. Como a soma do
módulo quadrado dos coeficientes da expansão de $\zeta$ vale $1$,
então $|f(\mathbf{n}_{a},\mathbf{n}_{b})| \leq 1$.

Por meio das Eqs.~(\ref{ai}-\ref{nb}) obtemos
\begin{eqnarray}
a'_{z} & = & \mathbf{a}_{i} \cdot \mathbf{n}_{a}, \\
b''_{z} & = & \mathbf{b}_{j} \cdot \mathbf{n}_{b}.
\end{eqnarray}
Assim, a Eq.~(\ref{fiadamae}) é reescrita como
\begin{equation}
E(\mathbf{a}_{i}, \mathbf{b}_{j})  =
|f(\mathbf{n}_{a},\mathbf{n}_{b})|(\mathbf{a}_{i} \cdot \mathbf{n}_{a})(\mathbf{b}_{j}
 \cdot \mathbf{n}_{b}), \label{fiadamae2}
\end{equation}
onde tomamos o módulo de $f(\mathbf{n}_{a},\mathbf{n}_{b})$
 para enfatizar que sempre podemos tê-lo positivo, simplesmente
 redefinindo os eixos $\mathbf{n}_{a}$ e $\mathbf{n}_{b}$.

Além disso, Eva pode mudar sua estratégia de medida para cada par
interceptado, bastando para isso alterar a orientação de $\mathbf{n}_{a}$
e $\mathbf{n}_{b}$. A função de correlação final se torna, pois,
\begin{equation}
E(\mathbf{a}_{i}, \mathbf{b}_{j})  =
\int_{}^{}\mathrm{d}\mathbf{n}_{a}\mathrm{d}\mathbf{n}_{b}
\varrho(\mathbf{n}_{a},\mathbf{n}_{b})(\mathbf{a}_{i}
\cdot \mathbf{n}_{a})(\mathbf{b}_{j} \cdot \mathbf{n}_{b}),
\label{Ecor}
\end{equation}
onde $\varrho(\mathbf{n}_{a},\mathbf{n}_{b})$ é probabilidade de
cada estratégia\footnote{Se $\varrho(\mathbf{n}_{a},\mathbf{n}_{b}) =
 |f(\mathbf{n}_{a},\mathbf{n}_{b})|\delta(\mathbf{n}_{a} -
 \mathbf{n'}_{a})\delta(\mathbf{n}_{b} - \mathbf{n'}_{b})$
 recuperamos a Eq.~(\ref{fiadamae2}). Ou seja, Eva fixou uma
 estratégia e a manteve para todas as medidas.} utilizada por Eva,
 i. e., $\int_{}^{}$$\mathrm{d}\mathbf{n}_{a}$$\mathrm{d}\mathbf{n}_{b}$
 $\varrho(\mathbf{n}_{a},\mathbf{n}_{b})$ $=$ $1$.

Finalmente, usando a Eq.~(\ref{Ecor}) a função $S$ pode ser assim escrita:
\begin{eqnarray}
S & = & \int_{}^{}\mathrm{d}\mathbf{n}_{a}\mathrm{d}\mathbf{n}_{b}
\varrho(\mathbf{n}_{a},\mathbf{n}_{b}) [ \nonumber \\
& & (\mathbf{a}_{1} \cdot \mathbf{n}_{a})(\mathbf{b}_{1}
\cdot \mathbf{n}_{b}) -  (\mathbf{a}_{1} \cdot \mathbf{n}_{a})(\mathbf{b}_{3}
\cdot \mathbf{n}_{b}) \nonumber \\
& & + (\mathbf{a}_{3} \cdot \mathbf{n}_{a})(\mathbf{b}_{1} \cdot \mathbf{n}_{b})
+ (\mathbf{a}_{3} \cdot \mathbf{n}_{a})(\mathbf{b}_{3} \cdot \mathbf{n}_{b})]
\nonumber \\
& = & \int_{}^{}\mathrm{d}\mathbf{n}_{a}\mathrm{d}\mathbf{n}_{b}
\varrho(\mathbf{n}_{a},\mathbf{n}_{b}) \{ \nonumber \\
& & (\mathbf{a}_{1} \cdot \mathbf{n}_{a}) [\mathbf{b}_{1}
\cdot \mathbf{n}_{b} - \mathbf{b}_{3} \cdot \mathbf{n}_{b}]
\nonumber \\
& & +  (\mathbf{a}_{3} \cdot \mathbf{n}_{a}) [\mathbf{b}_{1}
\cdot \mathbf{n}_{b} + \mathbf{b}_{3} \cdot \mathbf{n}_{b}] \}.
\label{S1}
\end{eqnarray}
Agora, como todos os vetores que aparecem na Eq.~(\ref{S1}) são
unitários, todos os produtos escalares têm módulo menor ou igual a $1$. Assim,
\begin{eqnarray}
|S| & \leq & \int_{}^{}\mathrm{d}\mathbf{n}_{a}\mathrm{d}\mathbf{n}_{b}
\varrho(\mathbf{n}_{a},\mathbf{n}_{b}) \{ \nonumber \\
& &|\mathbf{a}_{1} \cdot \mathbf{n}_{a}| |\mathbf{b}_{1} \cdot \mathbf{n}_{b}
 - \mathbf{b}_{3} \cdot \mathbf{n}_{b}|  \nonumber \\
& & +  |\mathbf{a}_{3} \cdot \mathbf{n}_{a}| |\mathbf{b}_{1} \cdot \mathbf{n}_{b}
 + \mathbf{b}_{3} \cdot \mathbf{n}_{b}| \}. \nonumber \\
& \leq  & \int_{}^{}\mathrm{d}\mathbf{n}_{a}\mathrm{d}\mathbf{n}_{b}
\varrho(\mathbf{n}_{a},\mathbf{n}_{b}) \{|\mathbf{b}_{1} \cdot \mathbf{n}_{b}
- \mathbf{b}_{3} \cdot \mathbf{n}_{b}| \nonumber \\
& & + |\mathbf{b}_{1} \cdot \mathbf{n}_{b} + \mathbf{b}_{3}
\cdot \mathbf{n}_{b}| \} \label{S2}
\end{eqnarray}
Analisando o termo entre chaves na Eq.~(\ref{S2}) vemos que ele é
 da forma $|x - y| + |x + y|$, onde $x = \mathbf{b}_{1} \cdot \mathbf{n}_{b}$
 e $y = \mathbf{b}_{3} \cdot \mathbf{n}_{b}$. Mas $|x - y|
 + |x + y| \leq |x| - |y| + |x| + |y| = 2|x|$, se $|x| > |y|$
 ou $|x - y| + |x + y| \leq |y| - |x| + |x| + |y| = 2|y|$,
 se $|x| < |y|$. Portanto, $|x - y| + |x + y| \leq \text{max}\{2|x|,2|y|\}$.
 E como $|x|\leq 1$ e $|y| \leq 1$ então $|x - y| + |x + y| \leq 2$.
 Usando este último resultado na Eq.~(\ref{S2}) e lembrando
  que $\varrho(\mathbf{n}_{a},\mathbf{n}_{b})$ está normalizada,
\begin{equation}
|S| \leq 2. \label{S3}
\end{equation}

A Eq.~(\ref{S3}) claramente mostra que qualquer interferência
feita por Eva nos pares de qbits que se dirigem até Alice e Bob
pode ser detectada por eles, pois nunca Eva conseguirá ao mesmo
tempo extrair alguma informação e reproduzir o valor $S =
-2\sqrt{2}$. Eva, no máximo\footnote{Se tivéssemos utilizado
explicitamente as orientações de $\mathbf{a}_{1}, \mathbf{a}_{3},
\mathbf{b}_{1}$ e $\mathbf{b}_{3}$, teríamos obtido um limite
superior ainda menor: $|S| \leq \sqrt{2}$.}, fará com que os
estados que cheguem a Alice e Bob alcancem $S = -2$ , não
importando a engenhosidade de sua estratégia. É neste sentido que
devemos considerar como garantido pelas leis da física o segredo
da chave criptográfica transmitida.

\section{BBM92}

Podemos simplificar ainda mais o protocolo anterior \cite{bbm92}.
Agora, ao invés de Alice e Bob orientarem seus detectores aleatoriamente
em três direções, eles necessitam apenas de duas direções. Eles orientam
seus polarizadores ou na direção $x$ ou na direção $y$. Note que ambas as
direções formam um ângulo de $90^{o}$.

Novamente, Alice e Bob anunciam publicamente a orientação de cada
 polarizador em cada medida. No entanto, eles não informam os
 resultados. Em seguida, eles descartam todas as medidas nas
 quais foram utilizadas orientações diferentes. São mantidos
  apenas os eventos cujos polarizadores foram orientados numa
  mesma direção. Se Eva não interferiu, toda medida onde
  ambos utilizaram uma mesma direção para seus polarizadores
  deve estar anticorrelacionada. Dessa forma, a grandeza
\begin{equation}
S = E(\mathbf{x},\mathbf{x}) + E(\mathbf{y},\mathbf{y}) = -2,
\label{siguala2}
\end{equation}
se Eva não interfere. Isso ocorre pois para o singleto
$E(\mathbf{x},\mathbf{x})= E(\mathbf{y},\mathbf{y}) = -1$.
Agora, se Eva interfere, mostramos na seção anterior que
\begin{equation}
E(\mathbf{a}_{i}, \mathbf{b}_{j})  =
\int_{}^{}\mathrm{d}\mathbf{n}_{a}\mathrm{d}\mathbf{n}_{b}
\varrho(\mathbf{n}_{a},\mathbf{n}_{b})(\mathbf{a}_{i} \cdot
\mathbf{n}_{a})(\mathbf{b}_{j} \cdot \mathbf{n}_{b}).
\end{equation}
Portanto,
\begin{eqnarray}
|S| & = &\left|\int\mathrm{d}\mathbf{n}_{a}\mathrm{d}\mathbf{n}_{b}
\rho(\mathbf{n}_{a},\mathbf{n}_{b}) \left[ (\mathbf{x}
\cdot \mathbf{n}_{a})(\mathbf{x}\cdot \mathbf{n}_{b}) \right. \right.
\nonumber \\
& & \left. \left.+ (\mathbf{y}\cdot \mathbf{n}_{a})(\mathbf{y}
\cdot \mathbf{n}_{b})  \right] \right| \nonumber \\
& = & \left|\int\mathrm{d}\mathbf{n}_{a}\mathrm{d}\mathbf{n}_{b}
\rho(\mathbf{n}_{a},\mathbf{n}_{b}) \left[ \mathbf{n}_{a_{x}}\mathbf{n}_{b_{x}}
+ \mathbf{n}_{a_{y}}\mathbf{n}_{b_{y}} \right]\right| \nonumber \\
& \leq & \int\mathrm{d}\mathbf{n}_{a}\mathrm{d}\mathbf{n}_{b}
\rho(\mathbf{n}_{a},\mathbf{n}_{b}) \left| \mathbf{n}_{a_{x}}\mathbf{n}_{b_{x}}
+ \mathbf{n}_{a_{y}}\mathbf{n}_{b_{y}} \right| \nonumber \\
& \leq & \int\mathrm{d}\mathbf{n}_{a}\mathrm{d}\mathbf{n}_{b}
\rho(\mathbf{n}_{a},\mathbf{n}_{b}) \left| \mathbf{n}_{a} \cdot \mathbf{n}_{b}
\right| \nonumber \\
& \leq & \int\mathrm{d}\mathbf{n}_{a}\mathrm{d}\mathbf{n}_{b}
\rho(\mathbf{n}_{a},\mathbf{n}_{b}) = 1. \label{siguala1}
\end{eqnarray}
A última desigualdade vem do fato de que $\mathbf{n}_{a}$ e
$\mathbf{n}_{b}$ são unitários. Observando a Eq.~(\ref{siguala1})
vemos que não há meios de Eva atingir o valor de $S$ dado pela
Eq.~(\ref{siguala2}). Assim, usando uma parte dos resultados
válidos, Alice e Bob podem calcular $S$. Se seu valor for dado
pela Eq.~(\ref{siguala2}), eles utilizam  a outra parte dos
resultados como chave. Caso o valor de $S$ seja diferente, eles
descartam todas as medidas e recomeçam o protocolo.

Na verdade, este protocolo é equivalente ao BB84. Para ver isso,
basta notar que as medidas de fótons com os polarizadores
orientados na direção $x$ ($y$) são equivalentes, no protocolo
BB84, aos fótons preparados por Alice na base A (B) e medidos por
Bob também nessa mesma base. A única diferença entre os dois
protocolos está na escolha dos números aleatórios a serem
transmitidos. No BB84 essa escolha é feita por Alice ao escolher
em qual base ela prepara seu fóton. Por outro lado, no BBM92 Alice
não tem mais essa liberdade. Nesse protocolo a seqüência é gerada
no momento em que Alice mede os seus fótons de cada singleto.

Para mostrar a segurança destes protocolos, pre\-ci\-sa\-mos
pro\-var que toda medida que não perturbe estados não-ortogonais
não fornece nenhuma informação sobre eles. Sejam $|u\rangle$ e
$|v\rangle$ estes estados, i. e., $\langle u|v\rangle \neq 0$.
Vamos representar pela transformação unitária $U$ a interação
entre os estados de Eva, nossa espiã, com os estados transmitidos
por Alice. O estado inicial de Eva é $|a\rangle$. Este estado é
bem geral. Eva pode usar quantos fótons julgar necessário. Assim,
para que Alice e Bob não percebam que Eva interferiu na
transmissão,
\begin{eqnarray}
U \left( |u\rangle |a\rangle \right) & = & |u\rangle |a'\rangle, \\
U \left( |v\rangle |a\rangle \right) & = & |v\rangle |a''\rangle.
\end{eqnarray}
Aqui, $|a'\rangle$ e $|a''\rangle$ são outros dois estados possíveis de Eva.
 Usando o fato de que $U$ é uma transformação unitária,
\begin{eqnarray}
  \langle a |\langle u |U^{\dagger}U  |v\rangle |a\rangle & = &
  \langle a'| \langle u|v\rangle |a''\rangle \nonumber \\
 \langle a|a\rangle \langle u |v\rangle & = & \langle u|v\rangle
 \langle a' |a''\rangle \nonumber \\
 \langle u |v\rangle & = & \langle u|v\rangle \langle a' |a''\rangle
 \nonumber \\
1 & = & \langle a' |a''\rangle. \label{aalinha}
\end{eqnarray}
A última igualdade decorre de $|u\rangle$ e $|v\rangle$ não serem
ortogonais. Mas a Eq.~(\ref{aalinha}) nos diz que $|a'\rangle$ e $|a''\rangle$
 são idênticos (estamos assumindo sempre estados normalizados). Dessa forma,
  qualquer medida que não perturbe os estados não-ortogonais não fornece
  nenhuma informação que permita a Eva distinguir entre eles. Provamos,
  assim, e de maneira bem geral, a segurança de qualquer protocolo que
  se utilize de estados não-ortogonais.

\section{B92}

Neste artigo \cite{b92} é demonstrada a possibilidade de se realizar
CQ utilizando apenas dois estado quânticos não-ortogonais (no protocolo
BB84 \cite{bb84} tínhamos quatro estados). Sua importância é mais conceitual
do que prática, pois esta proposta é difícil de ser implementada com as
tecnologias atuais. A motivação que levou Bennett a propor este protocolo
 é declarada no início de seu artigo: ``Na Ref. \cite{bbm92} a segurança
 dos sistemas de distribuição de chaves que não se utilizam de emaranhamento
  (\textit{BB84 é um exemplo}) advém do fato de que qualquer medida que
   não perturbe nenhum dos dois estados não-ortogonais também não
   fornece nenhuma informação que permita distinguir entre esses dois
   estados. Isto naturalmente sugere a possibilidade de que a
   distribuição de chaves possa ser realizada utilizando apenas
   \textit{dois} estados não-ortogonais...'' (tradução e ênfase nossas).

A demonstração de que apenas dois estados não-ortogonais são
suficientes é como se segue. Sejam $|A\rangle$ e $|B\rangle$
dois estados não-ortogonais ($\langle A|B\rangle \neq 0$) e sejam
$P_{A} = \mathcal{I} - |B\rangle\langle B|$ e $P_{B} = \mathcal{I}
 - |A\rangle\langle A|$, onde $\mathcal{I}$ é o operador identidade.
 $P_{A}$ e $P_{B}$ são operadores de projeção em espaços ortogonais
 a $|B\rangle$ e a $|A\rangle$, respectivamente (note os índices trocados).
 Dessa forma, $P_{A}$ aniquila $|B\rangle$ ($P_{A}|B\rangle = 0$),
 mas fornece um resultado positivo com probabilidade
 $Tr[P_{A}|A\rangle\langle A|] = 1-|\langle A|B \rangle|^{2}$
 quando é aplicado em $|A\rangle$. Resultado semelhante obtemos
 para $P_{B}$.  Alice e Bob devem combinar de antemão quais
 serão os estados $|A\rangle$ e $|B\rangle$ utilizados e qual
 corresponderá ao bit $0$ e qual ao bit $1$. $|A\rangle$ pode
 representar um fóton com polarização linear na direção dada
 por $\theta = \pi/2$ e $\phi = 0$ e $|B\rangle$ um fóton com
 orientação de polarização dada por $\theta = \pi/2$ e $\phi = \pi/4$.
 Veja Figs. \ref{esferica} e \ref{figbase}. Para começar a
 distribuição de chaves, Alice deve primeiramente escolher
 uma seqüência de bits e enviá-la a Bob codificando-a usando
 os estados $|A\rangle$ (bit $0$) e $|B\rangle$ (bit $1$).
 Bob, por sua vez, escolhe aleatoriamente para cada estado
 recebido de Alice qual medida realizará: $P_{A}$ ou $P_{B}$.
 Terminada a transmissão, Bob anuncia publicamente para
 quais bits ele obteve resultados positivos sem, no entanto,
 informar o tipo de  medida feita (se $P_{A}$ ou $P_{B}$).
 São estes bits que serão utilizados por Alice e Bob para
 obter a chave criptográfica. Como nos outros esquemas
 de CQ, alguns destes bits devem ser sacrificados para
 checar se Eva monitorou a comunicação. Assim, Bob
 publicamente informa que base utilizou para medir alguns
 de seus fótons. Se Eva não interferiu, todas as medidas nas quais Bob
 obteve um resultado positivo devem corresponder a duas únicas
 possíveis situações: 1) Alice enviou um estado $|A\rangle$ e Bob
 mediu $P_{A}$ ou 2) Alice enviou $|B\rangle$ e Bob mediu $P_{B}$.
 Caso ocorra um evento positivo para uma outra situação, Alice e Bob
 descartam seus bits pois Eva interferiu na transmissão. Se apenas estes
 dois eventos positivos ocorreram, eles têm certeza da segurança da chave,
 a qual é constituída pelos bits restantes.

Mostramos, agora, uma maneira de se implementar o protocolo B92.
Como dito no parágrafo anterior, supomos que $|A\rangle$ e
$|B\rangle$ representam estados de polarização linear de fótons.
As direções de polarização de ambos os fótons diferem de $45^{o}$.
Em coordenadas esféricas, para $|A\rangle$ temos $\phi=0$ e para
$|B\rangle$ temos $\phi = \pi/4$. Assim, a medida $P_{A}$ pode ser
feita utilizando um polarizador orientado na direção dada por
$\phi=3\pi/4$ (um polarizador transmite fótons com probabilidade
$\cos^{2}\alpha$, onde $\alpha$ é o ângulo entre as direções de
polarização do fóton e do polarizador). A medida de $P_{B}$ é
realizada por um polarizador orientado na direção $\phi=\pi/2$.
Veja Figs. \ref{esferica} e \ref{figbase}. Dessa forma, quando
Alice enviar um fóton representado pelo estado $|A\rangle$
($|B\rangle$) ele não será detectado por Bob quando ele fizer uma
medida $P_{B}$ ($P_{A}$), pois $P_{B}|A\rangle = 0$
($P_{A}|B\rangle = 0$). Este resultado já era esperado pois temos,
para estes casos, direções de polarização dos fótons ortogonais às
direções dos polarizadores. No entanto, quando Alice enviar o
estado $|A\rangle$ ($|B\rangle$) e Bob medir $P_{A}$ ($P_{B}$),
ele terá $50\%$ de chance de detectar um fóton (resultado
positivo). Isso ocorre pois o ângulo entre as direções de
polarização dos fótons e dos polarizadores é de $45^{o}$. Neste
esquema, se Alice enviou metade das vezes fótons descritos por
$|A\rangle$ e metade das vezes fótons descritos por $|B\rangle$ e
supondo que Bob mediu metade das vezes $P_{A}$ e metade das vezes
$P_{B}$, em apenas $25\%$ das vezes Bob obterá um resultado
positivo. Em outras palavras, metade das vezes ele não detectará
nada porque o eixo do seu polarizador será colocado numa direção
perpendicular à direção de oscilação dos fótons e, para a outra
metade, somente a metade fornecerá um resultado positivo, devido
ao eixo do polarizador estar a $45^{o}$ em relação ao eixo de
polarização dos fótons. E mais, se eles ainda sacrificarem metade
dos bits para testar a não interferência de Eva, eles obterão, no
final, como chave, uma seqüência aleatória de bits de tamanho
igual a $1/8$ do tamanho da seqüência original enviada por Alice.

Reiteramos que o protocolo B92 é conceitualmente importante pois
mostra a possibilidade de se fazer CQ utilizando apenas dois
estados não-ortogonais. Isso pode ajudar numa compreensão mais
intuitiva da CQ. Além disso, a nosso ver, é fácil imaginar uma
montagem experimental utilizando apenas polarizadores para
detectar os fótons, como feito no parágrafo anterior. Essa
simplicidade da montagem via polarizadores, o uso de apenas dois
estados e mais o fato de que a maioria dos estudantes tem uma boa
intuição física do que acontece quando um fóton passa por um
polarizador fazem do B92 um protocolo muito útil para se
apresentar a iniciantes no assunto.

\section{Discussão e Conclusão}

Neste artigo apresentamos de maneira acessível a estudantes de graduação
em Física os quatro protocolos de distribuição de chaves quânticas que
fundaram a área da Criptografia Quântica (CQ).

O primeiro deles, o protocolo BB84, é recomendado também como texto
introdutório a esse assunto. Devido a sua clareza e concisão, ainda
hoje consideramos a Ref. \cite{bb84} uma ótima opção para se introduzir CQ a
estudantes de Física.

O protocolo E91 requer um pouco mais de conhecimento do estudante.
Contudo, com um pouco de esforço e uma introdução às desigualdades de
Clauser, Horne, Shimony e Holt \cite{chsh} ele também pode ser ensinado
durante um curso de graduação em Física.

Os outros dois protocolos, por se tratarem de extensões e simplificações
dos protocolos anteriores, são facilmente entendidos por estudantes que
já dominaram o assunto dos primeiros dois protocolos.

Enfim, acreditamos na viabilidade de se ensinar CQ durante um
curso de graduação em Física. E mais, ensinar CQ pode ser também
extremamente vantajoso para convencer um público mais amplo da
importância da Mecânica Quântica.

\end{document}